\documentstyle[11pt,newpasp,twoside,psfig,wrapfig]{article}
\index{X-rays}
\index{Chandra}
\index{ASCA}
\index{XMM-Newton}
\index{jets}
\index{neutron stars!cooling}
\index{pulsars}
\index{pulsars!initial periods}
\index{pulsars!J0205+6449}
\index{pulsars!winds}
\index{pulsar wind nebulae!G130.7+3.1}
\index{pulsar wind nebulae!spectra}
\index{pulsar wind nebulae!spectral index variations}
\index{pulsar wind nebulae!wisps}
\index{supernova remnants!G130.7+3.1}
\index{supernova remnants!synchrotron emission}

\newcommand{\chandra}{{\em Chandra}}

\def\edcomment#1{\iffalse\marginpar{\raggedright\sl#1\/}\else\relax\fi}
\reversemarginpar

\begin{document}
\title{High resolution X-ray observations of 3C~58}
 \author{Patrick Slane}
\affil{Harvard-Smithsonian Center for Astrophysics, 
60 Garden Street,
Cambridge, MA 02138, USA}

\begin{abstract}
 As the presumed remnant of SN1181, 3C~58 houses one of the youngest known
neutron stars in the Galaxy. The properties of this young pulsar and its
associated wind nebula differ considerably from those of the Crab, and may well
offer a more typical example of the endpoint of massive star collapse. 
High resolution X-ray
studies reveal structures in the inner nebula that may be associated
with the pulsar wind termination shock, a jet that may be aligned with the
rotation axis, and other regions of enhanced emission. Spectral variations in
the PWN are consistent with the expected evolution of the postshock flow, 
and complex loops of emission are seen in the nebula interior.
Limits on the NS surface temperature
fall below standard cooling models, indicating that some more rapid neutrino
cooling process is required. The outer regions of 3C~58 show thermal emission
with enhanced levels of Ne, indicative of shocked ejecta bounding the PWN.
\end{abstract}

\section{Introduction}

3C~58 (G130.7+3.1) is a Crab-like nebula believed to be associated 
with the historical supernova SN~1181. X-ray observations reveal
reveal a compact source at the center, and a power
law spectrum whose index steepens with radius, characteristic of particle
injection from a central pulsar. 3C~58 appears to be quite different from the
Crab Nebula in many respects, however. The nebula is
larger, and the radio luminosity is ten times lower than that
of the Crab. The X-ray luminosity is nearly 2000 times lower, perhaps suggesting
that any central pulsar has undergone a significant reduction
in its injection power. However,
recent \chandra\ HRC observations (Murray et al.\ 2002) have revealed 
PSR~J0205+6449, the second most energetic pulsar known in the
Galaxy, powering the nebula; the
characteristic age is $\tau \equiv P/2\dot P \approx 5 \times 10^3$~yr which,
assuming a true age of $\sim 820$~yr based on the historical association, 
implies an initial spin period of $P_0 \approx 60$~ms --- much
longer than that inferred for the Crab pulsar.

The emission from the center-most region of 3C~58 
is extended, with elongation in the N-S direction, perpendicular to
the long axis of the main nebula. The pulsar resides at the center of
this compact core nebula, and a jet-like feature protrudes westward.  
At the western edge resides a narrow
radio wisp discovered by Frail \& Moffett (1993),
who suggested that this lies along the pulsar wind termination
shock. Using the measured value of
$\dot E = 10^{37.6}{\rm\ erg\ s}^{-1}$ 
and the pressure determined for
the nebula interior, the expected radius of the termination shock 
is $r_w \approx 12^{\prime\prime}$, in good agreement with the
north-south extent of the core nebula in 3C~58 (Slane et al. 2002).
This suggests a toroidal structure with an inclination angle of 
$\sim 70^\circ,$ and with a projected pulsar spin axis in the E-W direction. 
The elongated shape of the entire PWN is consistent
with a pulsar axis oriented in the E-W direction (Begelman \& Li 1992, 
van der Swaluw 2003).

While unresolved from the compact surrounding nebula, limits on the
X-ray flux from PSR~J0205+6449 provide important constraints on 
models for NS cooling.
The upper limit to the thermal emission from the NS surface falls well below
predictions from standard cooling models and appears to require the
presence of some rapid cooling mechanism (Slane, Helfand, \& Murray 2002).
Such rapid cooling may imply an above-average NS mass, for which an increased
proton fraction in the interior allows the direct Urca process to proceed
(Yakovlev et al. 2002), or may indicate the presence of pion condensates
in the NS interior, which also lead to rapid cooling (Tsuruta et al. 2003).

\begin{figure}[t]
\centerline{\psfig{file=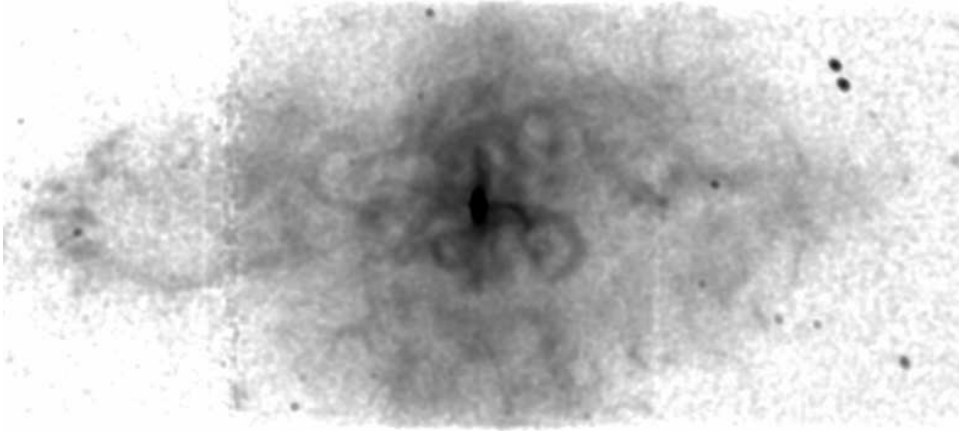,width=5in}}
\caption{
350 ks X-ray image of 3C~58 taken with the \chandra\ ACIS detector. The
compact region in the center is extended, and reveals a jet-like feature
protruding to the west. On larger scales, complex loop structures are observed
throughout the PWN. 
}
\end{figure}

\begin{figure}[t]
\centerline{\psfig{file=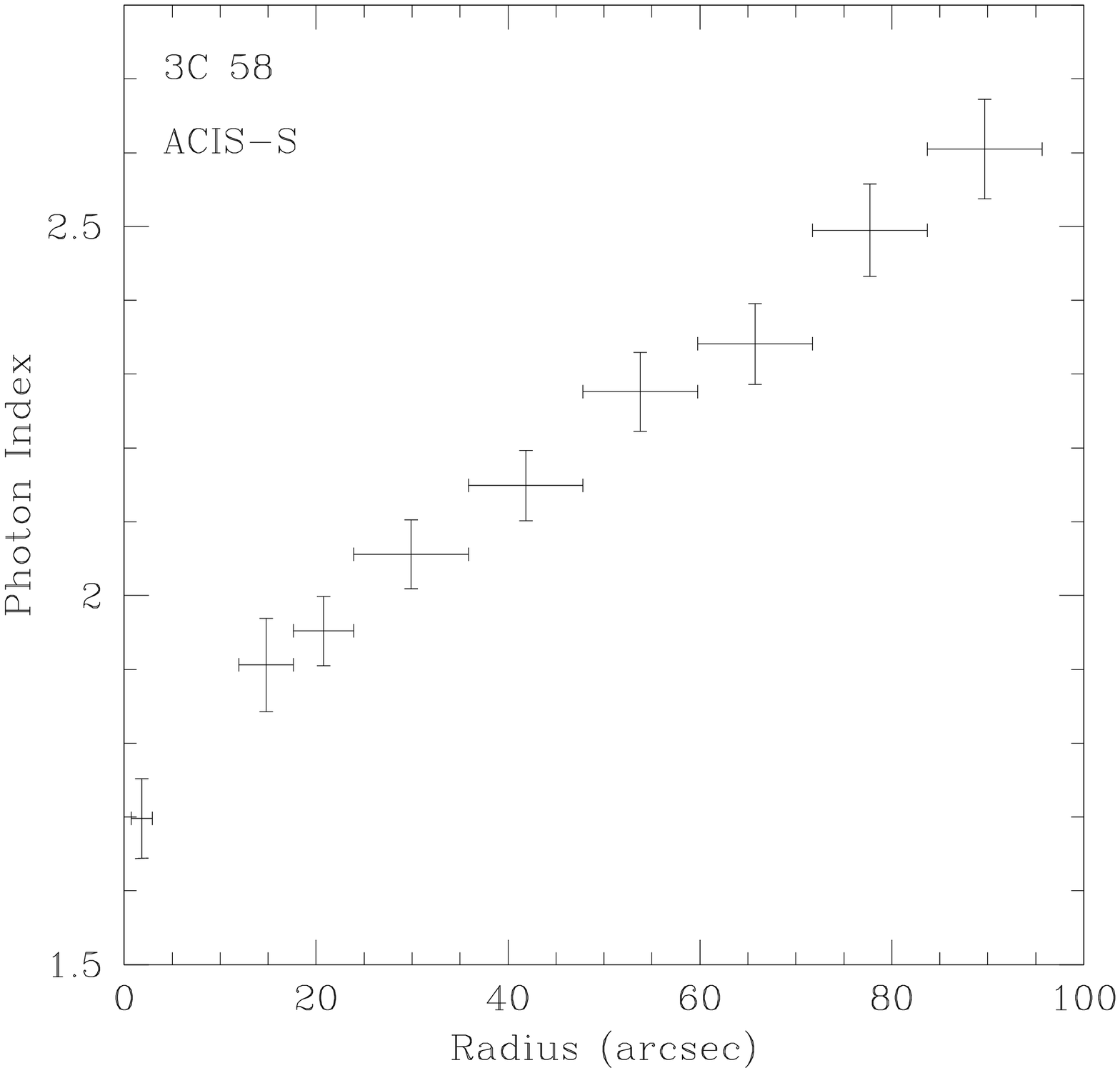,width=2.5in}
\psfig{file=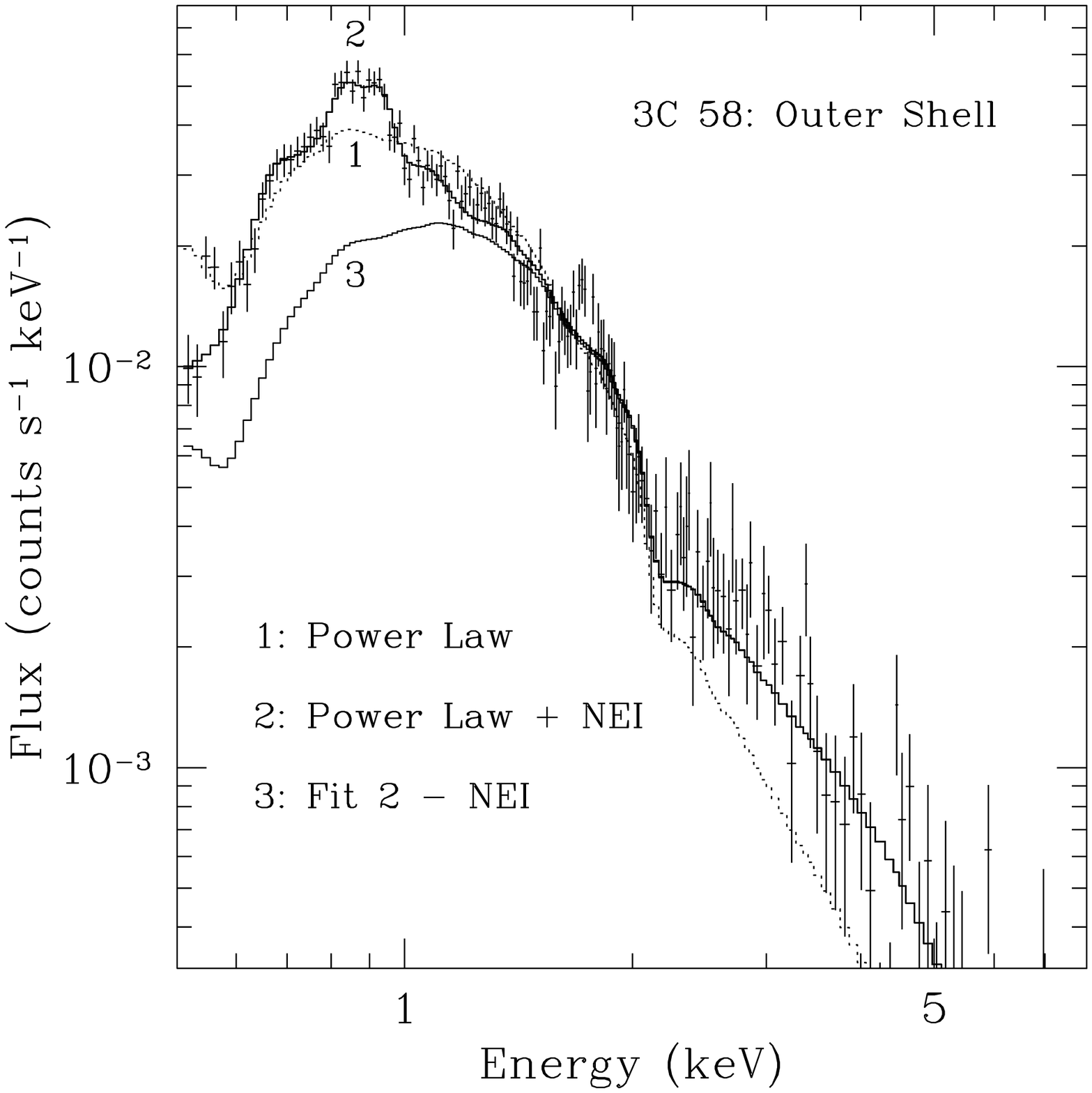,width=2.5in}}
\caption{
Left: Variation of the spectral index with radius in 3C 58.
Right: Spectrum of the outermost regions showing the best-fit
power law model (1); a power law
+ thermal model (2) that requires an overabundance of Ne; and the contribution
from only the power law (3) from the best-fit two component model.
}
\end{figure}

\section{New Observations of 3C~58}
3C~38 was recently observed for 350~ks with the ACIS detector onboard 
the \chandra\ Observatory.
The X-ray image, shown in Figure 1 (Slane et al. 2003, in preparation)
and described in more detail below, reveals a complex structure consisting of 
loops, elongated features, and broad diffuse emission.
The observation was divided into three segments to accommodate interruptions 
by the radiation zones in the \chandra\ orbit. An error in on-board commanding 
resulted in the final segment being carried out with the spacecraft dither 
turned off for the full 170~ks duration, resulting in no exposure in regions 
corresponding to gaps between the CCD chips or to bad column regions in the 
CCDs for this segment; these result in faint vertical features evident in
Figure 1.

\subsection{Spatial and Spectral Structure in 3C 58}

X-ray spectra were extracted from various regions of 3C~58, with corresponding
background spectra taken from a region of the CCD beyond the extent of the
nebula. The emission is well described by a power law model, 
typical of synchrotron emission. The spectral index varies throughout the PWN, 
however.  We find, in particular, that the average spectral index increases 
with radius (Figure 2, left), as expected for synchrotron losses as the 
electrons diffuse from the center, consistent with the results 
based on studies of 3C~58 with {\em ASCA} (Torii et al. 2000) and 
{\em XMM-Newton} (Bocchino et al. 2001).
Here we have forced the column density to be the same for each emission
region, but have left the overall value as a free parameter in the spectral
fit. The
best-fit value for the column density is $N_H = (3.69 \pm 0.06) \times
10^{21}{\rm\ cm}^ {-2}$, and we use this value when investigating the
emission from discrete regions from throughout the nebula.

For the jet-like region protruding from the pulsar, we extracted spectra
from three distinct regions extending across the feature. Fixing
the column density for each region at that derived above, we find a
spectral index of $\sim 2.0$ with no significant evidence for variations
from region to region.

As noted above, one of the remarkable features revealed by the deep
ACIS observation of 3C~58 is the presence of loop-like structures throughout
the nebula, suggestive of magnetic structures in which the synchrotron emission
is enhanced. We have investigated the spectra from 
individual loops, and compared these with the emission in the ``void'' regions
interior to the loops to search for variations that might be
expected if the synchrotron-emitting particles age as they diffuse from the
loop regions. While we find spectral index differences between loops in the
interior and exterior regions of the nebula, the void regions have spectra
similar to those of the corresponding loops. The jet-like feature
extending eastward from the pulsar appears to merge with a loop structure
to the southeast, possibly suggesting that the former is not actually
a jet. However, the spectrum of this feature is flatter than
for the adjacent loop. While this does not confirm the jet notion, it does
indicate that this is a distinct feature.

The X-ray spectrum from the outermost region of 3C~58 (Figure 2, right) is 
not well fit by a simple power law. Residual emission below $\sim 1$~keV 
requires an additional thermal model with $kT \sim 0.25$~keV and roughly a 
factor of 3 overabundance of Ne. This is similar to results reported by
Bocchino et al. (2001) based on {\em XMM-Newton} observations, and represents
confirmation of the long-sought thermal shell bounding 3C~58. The density
of the thermal plasma is of order $n \sim 0.05 {\rm\ cm}^{-2}$, corresponding
to an observed mass of $\sim 0.06 M_\odot$, although these values are quite
uncertain due to the poorly defined geometry of the thermal emission region.
The presence of enhanced Ne suggests that the thermal emission is associated
with stellar ejecta. However, if associated with reverse-shocked ejecta, this 
would imply the presence of a considerable swept-up ISM component outside the
PWN, which is not observed. If, instead, the emission is from ejecta swept up
the the PWN itself, the size, age, and spindown power for the system imply
a relatively small total ejecta mass (Chevalier 2003). 

\section{Conclusions}
3C~58 is a young PWN powered by an energetic pulsar. The compact interior 
shows evidence of structure from the pulsar wind termination
shock. Limits on thermal emission from the neutron star itself require rapid
cooling mechanisms such as the direct Urca process or the presence of exotic
particles in the NS interior. The overall structure of the PWN itself is
dominated by a significant east-west elongation, consistent with a pulsar
rotation axis in this same direction, along with complex loop-like structures
that may represent magnetic structures in the nebula. Thermal emission from
the outer regions of the nebula appear to correspond to ejecta-rich material
that is bounding the PWN.

\acknowledgments
This work was supported by NASA Contract NAS8-39073 and Grant GO0-1117A.

\end{document}